\documentclass[11pt,a4paper,dvips]{article}
\usepackage{a4p}
\usepackage{cite,mcite}
\usepackage{epsfig}
\usepackage{graphicx}
\usepackage{subfigure}
\usepackage{atlasphysics}
\usepackage{atlas_title,ifthen}
\usepackage{mathptmx}
\usepackage{helvet}
\newlength{\capindent}
\newlength{\capwidth}
\newlength{\figwidth}
\newcommand{\icaption}[2][!*!,!]{\hspace*{\capindent}%
  \begin{minipage}{\capwidth}
    \ifthenelse{\equal{#1}{!*!,!}}%
      {\caption{#2}}%
      {\caption[#1]{#2}}
      \vspace*{3mm}
  \end{minipage}}
\def\etmiss{\slashchar{E}_T}
 
 \def\slashchar#1{\setbox0=\hbox{$#1$}           
    \dimen0=\wd0                                 
    \setbox1=\hbox{/} \dimen1=\wd1               
    \ifdim\dimen0>\dimen1                        
       \rlap{\hbox to \dimen0{\hfil/\hfil}}      
       #1                                        
    \else                                        
       \rlap{\hbox to \dimen1{\hfil$#1$\hfil}}   
       /                                         
    \fi}

\begin{document}
\begin{titlepage}
\vspace*{-6mm}
\includegraphics[width=3cm]{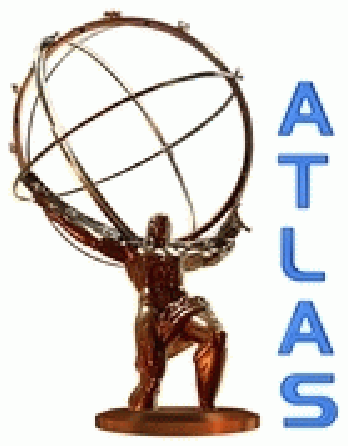} \hfill
\begin{minipage}[b]{7cm}
\begin{center}
\mbox{\Huge \bf ATLAS NOTE}
\end{center}
\begin{center}
\mydocversion
\end{center}
\begin{center}
\end{center}
\end{minipage}
\hfill \includegraphics[width=3cm]{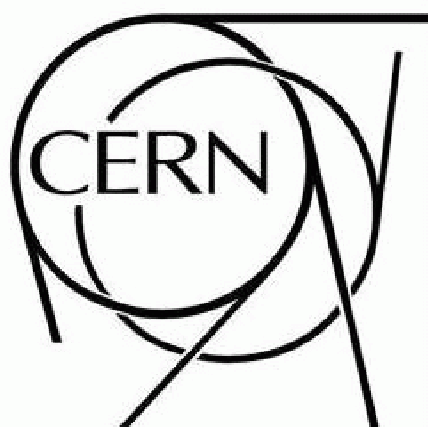}

\title{ Parton Densities at the LHC}
\author{Alessandro Tricoli (RAL)}
%
%
\begin{abstract}
This contribution to the Italian ``Workshop sui Monte Carlo, la Fisica e le Simulazioni a LHC'', held at LNF, Frascati, in February, May and October 2006, summarises the status of parton density functions (PDF's) and the impact of their uncertainties on the LHC physics. Emphasis is given to methods of contraining PDF's using LHC data. Moreover, the advantages of the so-called PDF reweighting technique, which enables to quickly estimate PDF uncertainties with Monte Carlo events, are also presented.   

\end{abstract}
\end{titlepage}
%
\section{Introduction}

The start up of the LHC machine is now imminent and theorists and experimentalists are converging their efforts to enhance the LHC discovery potential. 
This implies
 minimising theoretical and experimental uncertainties.
Among the theoretical uncertainties the knowledge of the proton structure plays a major role: the accurate evaluation of parton density functions (PDF's) is vital to provide reliable predictions of new physics signals (i.e. Higgs, Supersymmetry, Extra Dimensions etc.) and their background cross sections at the LHC. 
As shown in the contribution by C. Mariotti, E. Migliore and P. Nason
, at hadron colliders the inclusive cross section for hard production processes is the convolution of the cross section at parton level, calculable at fixed order in perturbation theory, and the parton densities
of the two interacting partons.

\begin{figure}[!t]
\begin{center}
\includegraphics[scale=0.35]{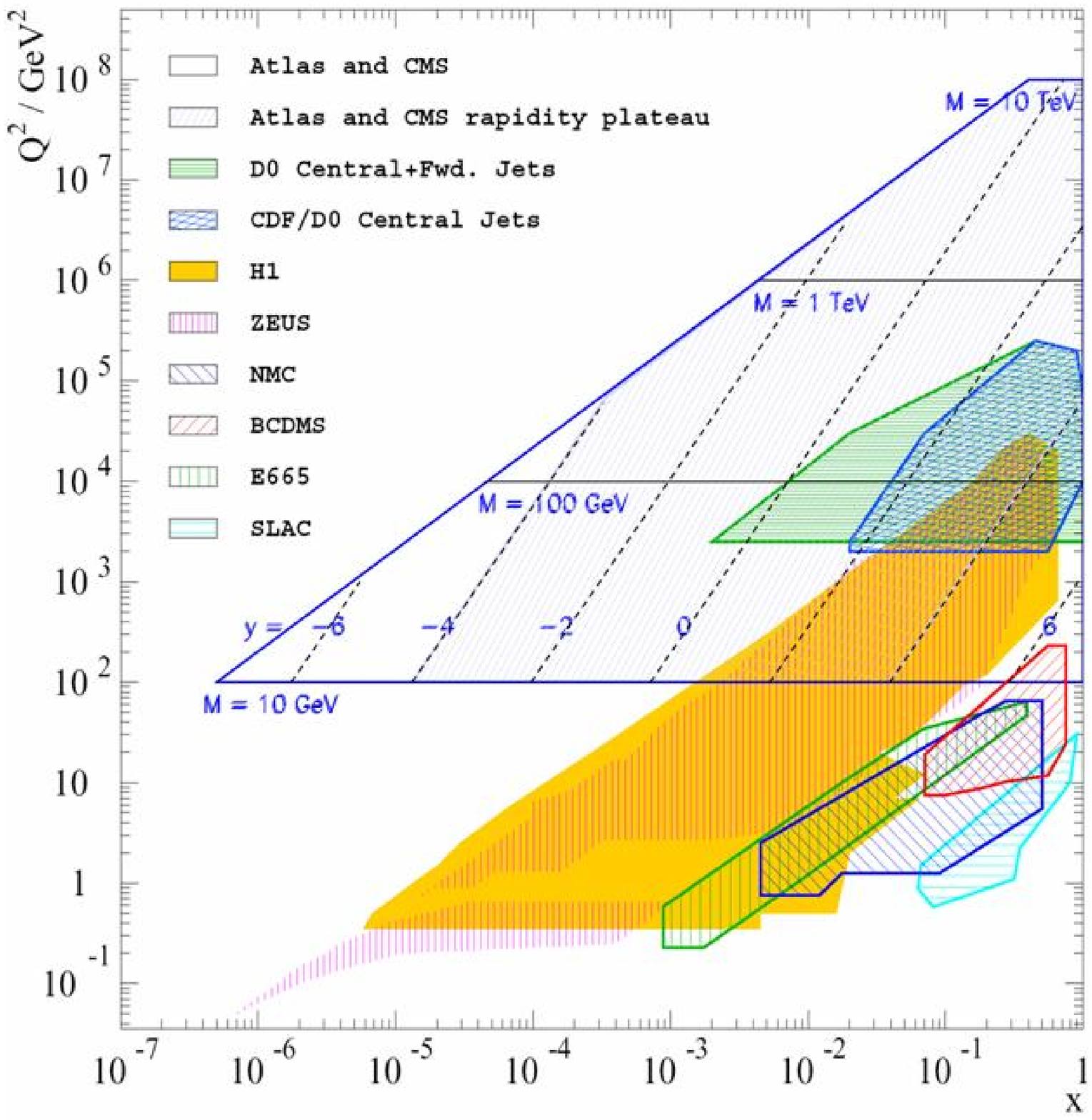}
\end{center}
\caption[LHC Kinematic Regime]{\label{fig:LHCKinRegime}}{\emph{The $Q^2$-$x$ kinematic plane for the LHC and previous experiments, showing the mass ($M$) and rapidity ($y$) dependence.}}
\end{figure}

Our knowledge of the proton structure is improving fast thanks to more experimental data being available and thanks to more precise and sophisticated theoretical calculations: PDF's are nowadays available up to the next-to-next-to leading order ({\em NNLO}) in perturbative QCD and in recent years they have been also providing uncertainties which take into account experimental systematic errors and the correlations between data points that enter the global fits.
Despite the great improvement on PDF's in recent years, their uncertainty dominates many cross section calculations for the LHC. As visible in fig.~\ref{fig:LHCKinRegime}, the LHC will probe kinematic regions in $x$ (parton momentum fraction) and $Q^2$ (hard scattering scale) never explored before, such as the $very ~high$-$Q^2$ and the $very ~low$-$x$ regions. At low-$x$ the current theoretical formalism (DGLAP) is at the edge of its supposed applicability. 
For the production of $Z$ and $W$ bosons the participating partons have small momentum fractions at central rapidity, $x \sim 10^{-3}$, and in the whole measurable rapidity region, $|y|<2.5$, they are within the range $10^{-4}<x<0.1$. Thus, at the electro-weak scale the theoretical predictions for the LHC cross sections are dominated by low-$x$ PDF uncertainty. At the ${\rm TeV}$ scale, where we expect new physics, the interacting partons have higher momentum fractions and very high $Q^2$ ($\geq 10^6~{\rm GeV^2}$).
Thus, at the ${\rm TeV}$ scale the cross section predictions are dominated by high-$x$ PDF uncertainty and rely on the extrapolation of the DGLAP equations. 
In both kinematic regimes the gluon density, which is in most regions the less well constrained density function, plays a major part: at low $x$ the gluon density dominates the quark and anti-quark densities, at high $Q^2$ the interacting partons get an important contribution from the sea, which is driven by the gluon density, via the $g \rightarrow q\bar{q}$ splitting process. For a review on hard interactions of quarks and gluons at the LHC refer to~\cite{LHCPrimer}.

Past and running experiments, such as HERA, have been providing vital information to improve our knowledge of the parton densities, however the broad kinematic region of the LHC forces (and offers a unique opportunity to) ATLAS and CMS experiments to use their own data to constrain the parton densities, in particular the gluon, in the kinematic regions where they are not sufficiently well determined.
In section~\ref{sec:PDFconstraints} it will be shown that significant improvement on PDF fits can be made with LHC data.

\section{Global fits and error analysis}

Perturbative QCD provides the evolution equations for the parton densities, DGLAP equations, but does not provide us with their analytic forms as function of $x$.
The most common approach to extrapolate PDF's as function of $x$ and $Q^2$ consists in solving the DGLAP equations by parameterising the parton densities $q_i(x)$ at a fixed scale $Q^2_0= 1 - 7 ~{\rm GeV^2}$, applying assumptions and constraints derived from theory and measurements. Then, with the DGLAP equations, we numerically extrapolate the values of $q_i(x,Q^2)$ to different values of $Q^2$ and a global fit of experimental data is performed.
For valence quarks the parameterisations have usually this behaviour $q_V \approx x^\lambda (1-x)^\eta $, whereas for the gluon and sea quarks they are of this kind $q_S(g) \approx x^{- \lambda} (1-x)^\eta $.
However there is no unanimous agreement on the parametric functions to use and on the number of free parameters. For a review refer to~\cite{Devenish_Cooper-Sarkar}.

Different regions in the $x, Q^2$ plane and also different partonic components are probed by the available world experimental data. These include DIS data from fixed target experiments and HERA, Drell-Yan data, inclusive jet production and $W$ charge asymmetry from Tevatron.

There are various groups who are fitting the proton structure function data, among them CTEQ and MRST. Recent PDF sets include in their analyses up-to-date experimental data and attempt to provide coherent estimates of the uncertainties, including experimental correlated systematic errors. The differences between these PDF sets can be summarised in three categories: different choices of input data sets, different theoretical model assumptions and different error analyses.

There are many sources of uncertainty which contribute to a global fit uncertainty. These can have experimental and theoretical origins. The former are related to the data errors which enter the fit, the latter are due to the model uncertainties of the theoretical framework.
The theoretical uncertainties concern both the non-perturbative (parameterisations) and perturbative parts of the calculations: assumptions imposed to limit the number of free parameters, higher order truncations in the DGLAP formalism etc.

The treatment of the experimental uncertainties, especially the correlated systematic uncertainties, is a complex subject which is partly still under debate. A modified version of the standard $\chi^2$ method is used to take into account non-Gaussian systematic errors and their correlations: $\chi^2\rightarrow \tilde\chi^2+\Delta T^2$, where $\Delta T$ is the so-called ``tolerance'', a complicated mathematical expression that includes correlated systematic terms~\cite{Devenish_Cooper-Sarkar}.
There are then two methods to compute the central values of the theoretical PDF parameters and their uncertainties: the {\em offset} and the {\em Hessian} method. In the offset method the correlated systematic errors affect only the determination of the PDF uncertainty, not the best fit. This method is used for ZEUS PDF's.
Conversely in the Hessian method, used by CTEQ and MRST groups, the collective effect of the correlated systematic errors has also an impact on the best fit.

\begin{figure}[!tb]
  \begin{minipage}[t]{.40\textwidth} 
  	\begin{center}  
	\includegraphics[width=7.5cm, height=8.3cm]{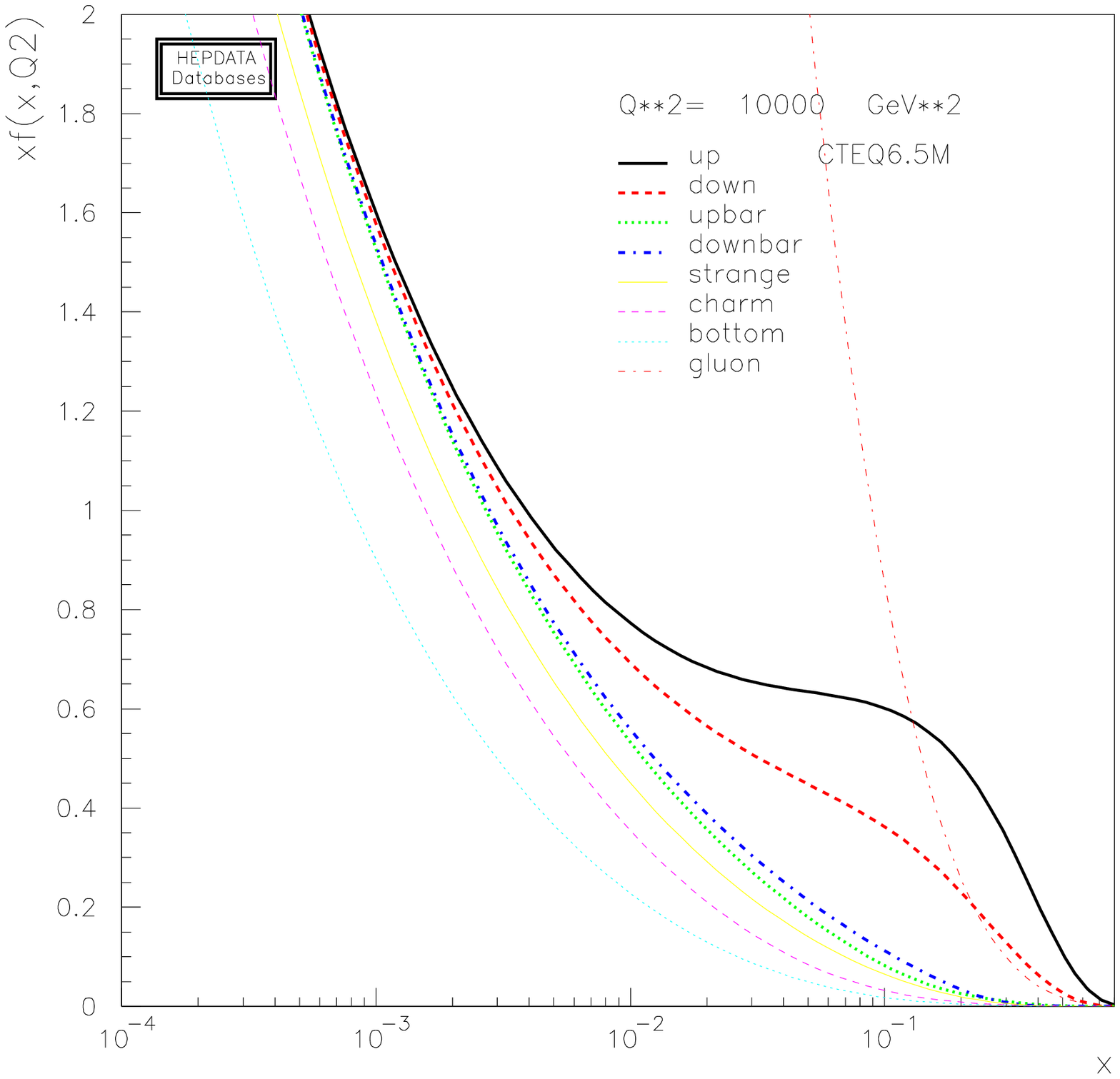}    
	\end{center}
  \end{minipage}
  \hspace{10mm}
  \begin{minipage}[t]{.40\textwidth}
    \begin{center}
      \includegraphics[width=7.0cm, height=8.cm]{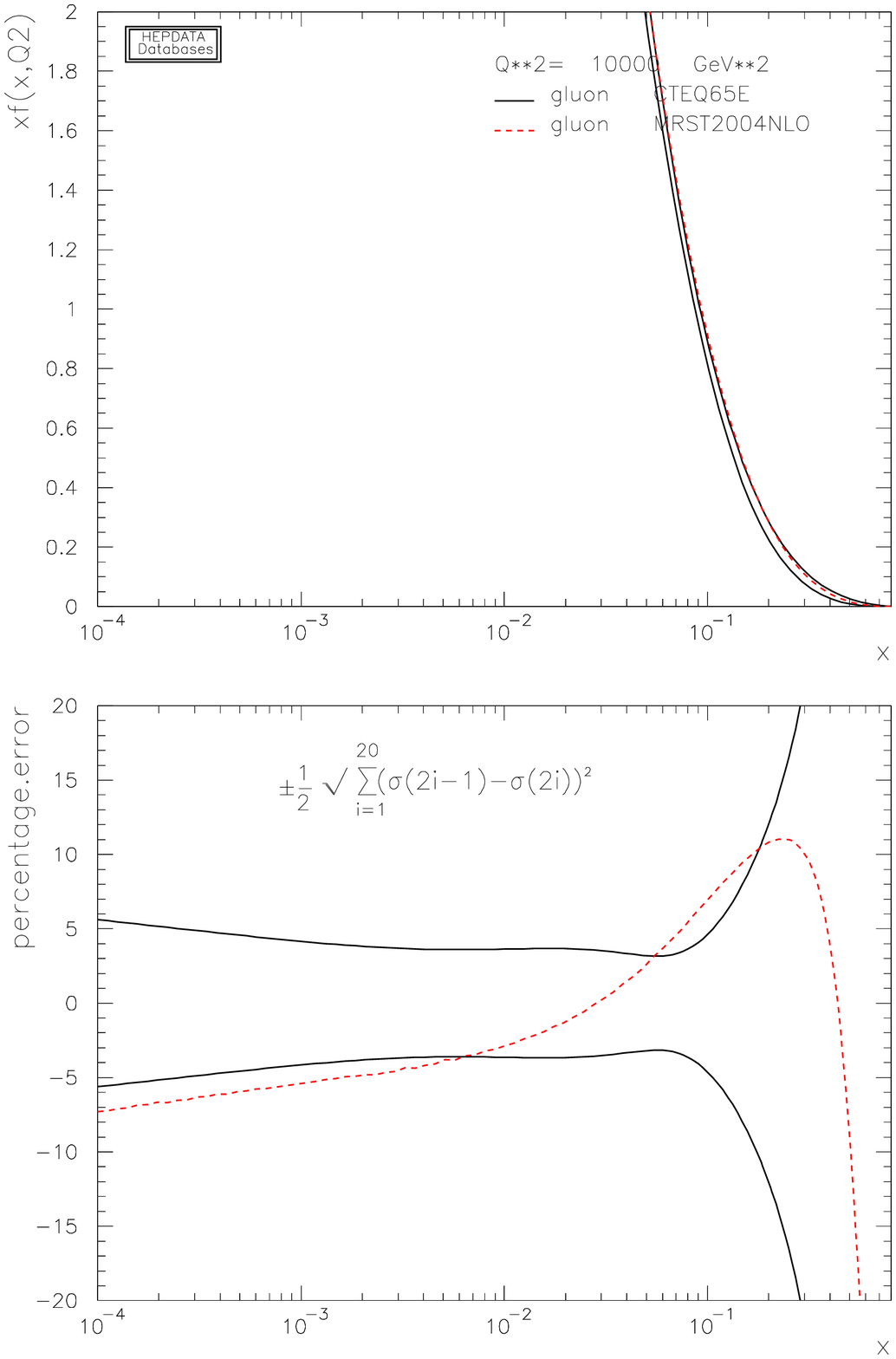}  
    \end{center}
  \end{minipage}
  \caption[CTEQ65 and MRST2004NLO PDF]{\label{fig:CTEQ65vsMRST2004NLO}}{\emph{Left: CTEQ6.5M set at $Q^2\sim M_W^2$. Right: comparison between CTEQ6.5M (black) and MRST2004NLO (red) gluon PDF's and their uncertainties.}} 
\end{figure}

For both, the offset and the Hessian methods, the PDF uncertainty is conventionally computed along the eigenvectors of the diagonalised covariance or Hessian matrices.
The number of eigenvectors corresponds to the number of free parameters in the parton density parameterisations.
Contemporary PDF sets provide a central value PDF set, corresponding to the best data fit, and two PDF sets for each uncertainty eigenvector, giving the upper and lower limit on the uncertainty. Given a PDF set, the upper limit of the PDF uncertainty is calculated for a physical observable by adding in quadrature the upward displacement eigenvectors, whereas the lower limit by adding in quadrature the downward displacement eigenvectors. 
MRST group has chosen 15 free parameters, leading to 30 error sets; CTEQ6 has 20 free parameters and 40 error sets.
Fig.~\ref{fig:CTEQ65vsMRST2004NLO} shows CTEQ6.5 fit for all parton densities at the scale $Q^2\sim M_W^2$ and its gluon uncertainty compared to the MRST2004NLO gluon best fit.

\section{Impact of PDF uncertainty on LHC physics}
The experience from previous experiments teaches that the PDF uncertainties must be properly taken into account or features of the SM physics can be misinterpreted as evidence of new physics. For example an unexplained discrepancy between data and theory was originally found in the Tevatron Run-I jet data, which was subsequently reabsorbed within the theoretical uncertainty when a more accurate PDF error analysis was performed.

G. Polesello's contribution on inclusive jet cross-section has shown that the PDF uncertainty is dominating for high $E_T$ jets over the renormalisation/factorisation scale and the experimental energy scale uncertainties: $10\%$ at $1~{\rm TeV}$, $25\%$ at $2~{\rm TeV}$, $60\%$ at $5~{\rm TeV}$.

\subsubsection{Extra dimensions}
In extra dimensions models, if the compactification scale $M_C$ is about few TeV\footnote{In this context the compactification scale is defined as $M_C = 1/R_C$ where $R_C$ is the compactification radius of the extra dimensions on a hypersphere.}, it is possible to observe the production of gravitons and Kaluza Klein (KK) excitations at the LHC. If gauge bosons can propagate in the extra dimensions, we also expect a violation of the SM logarithmic behaviour of the running couplings.
In this scenario, if we consider the CTEQ6M PDF uncertainty on the di-jet cross-section, we see the extra dimensions prediction being absorbed within the SM prediction zone: the high-$x$ gluon uncertainty can cause a decrease of the discovery reach from $M_C = 5~(10)~{\rm TeV}$ to $M_C < 2~(3)~{\rm TeV}$, depending on the number of extra dimensions~\cite{PDFError_ED}.

\subsubsection{Higgs}
The accurate measurements of the Higgs production cross sections and decay branching ratios are crucial to explore all Higgs boson fundamental properties. At the same time, we need very precise estimates of the various theoretical uncertainties.

It is found that the PDF uncertainty can be of the same order of magnitude or even higher than the other theoretical uncertainties. In fact the perturbative calculations of Higgs production cross section are becoming more stable as higher orders are included, leaving the PDF uncertainty as one of the largest contributions to the total theoretical uncertainty.
For example for the dominant Higgs production channel, $gg\rightarrow H$, the PDF uncertainty on $gg$ luminosity, can be larger than the factorisation and renormalisation scale uncertainty: in fact the differences in the $gg$ luminosity prediction between MRST2002 and Alekhin2002 can be higher than $10\%$ for low Higgs mass scenarios. 
Furthermore, studying the effect of three different PDF sets (i.e. CTEQ6M, MRST2001E and Alekhin2002) with their quoted uncertainties, on various Higgs productions channels, we see that the PDF uncertainty can be of the order of $\sim 10-15\%$ on the production cross-section~\cite{PDFError_Higgs}. 

\subsubsection{High mass Drell-Yan}

Several new physics models predict events with two charged leptons originating from the decay of a massive object. A peak in the $d\sigma/dM$ distribution is a clean signature of a new resonance: the identification and reconstruction of high-mass di-lepton final states can be done with high efficiency and the SM background can be small.
However the shape and normalisation of the predicted observable distributions depend on PDF and its uncertainty.

\begin{figure}[!t]
 \hspace{-10mm} 
  \begin{minipage}[t]{.40\textwidth} 
  	\begin{center} 
       \includegraphics[width=11cm, height=6.5cm]{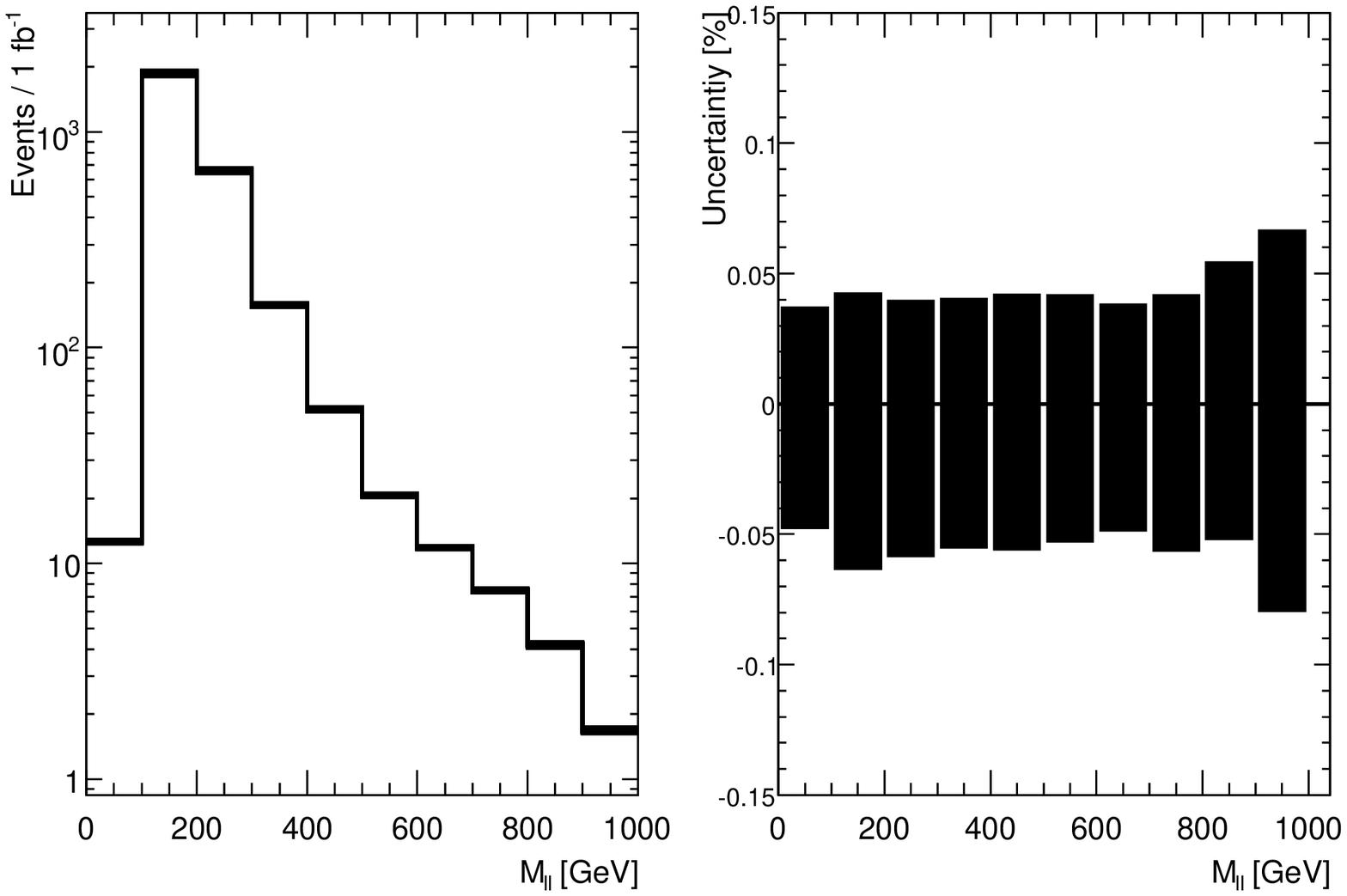}  
	\end{center}
  \end{minipage}
  \hspace{45mm}
  \begin{minipage}[t]{.40\textwidth}
    \begin{center}  
      \includegraphics[width=6.5cm, height=6.5cm, clip=true, viewport=280pt 0pt 567pt 414pt]{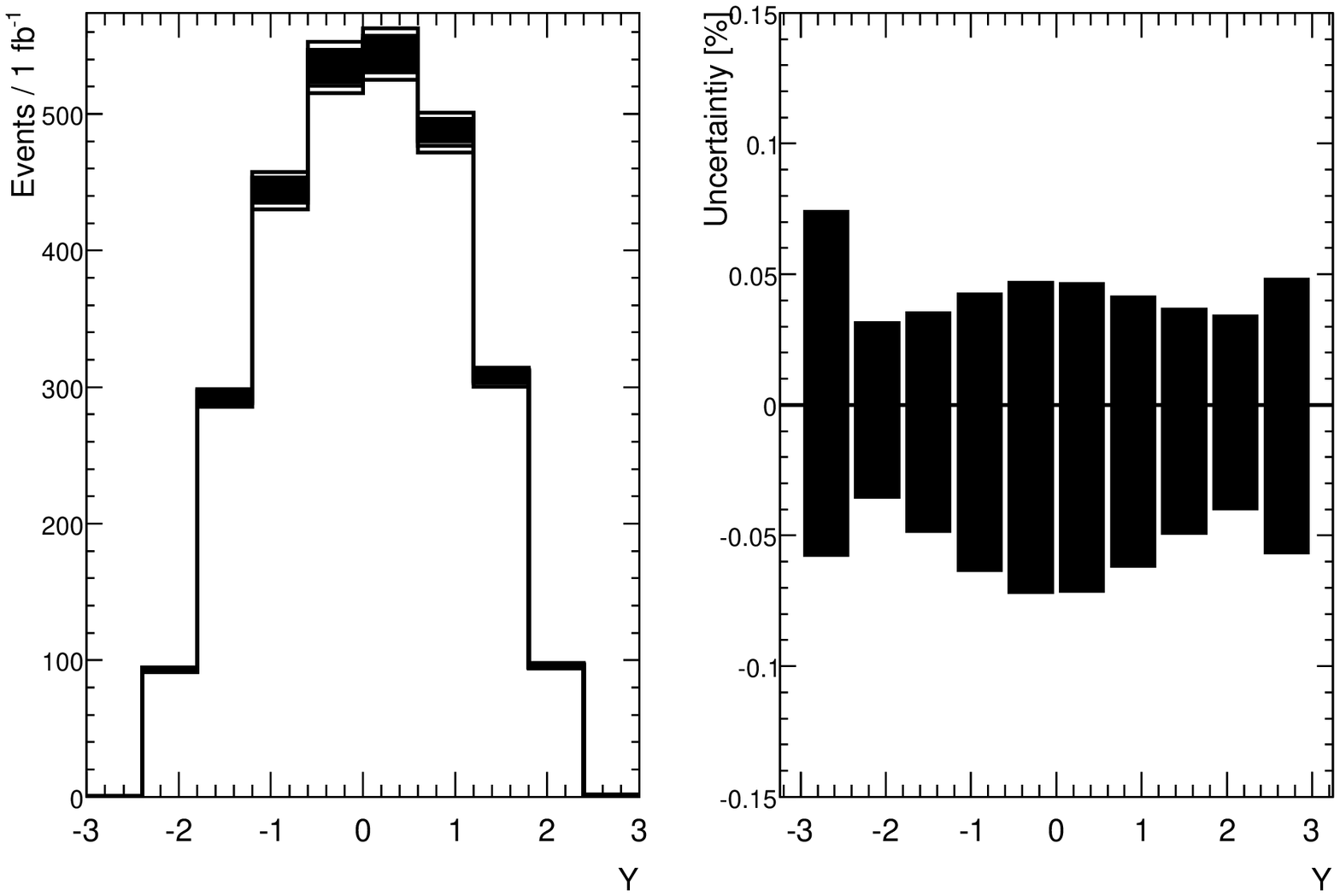}
    \end{center}
  \end{minipage}
  \caption[High Mass Drell-Yan]{\label{fig:HighMassDilepton}}{\emph{CTEQ6.1 uncertainty on distributions of the high-mass di-electron $M_{ll}$ (left and centre) and rapidity $y$ (right). Herwig+Jimmy generation and ATLAS full simulation~\cite{Florian}.
N.B.: the drop in the low $M_{ll}$ spectrum is an artifact of the event selection in the Monte Carlo.
}} 
\end{figure}

In fig.~\ref{fig:HighMassDilepton} we see the total CTEQ6.1 uncertainty on the distributions of the reconstructed rapidity $y$ and invariant mass $M_{ll}$ of the lepton pair: 40 CTEQ error sets have been accounted for, applying the PDF reweighting technique (see sec.~\ref{sec:PDFreweighting}). The uncertainty is in the range $4-7\%$ on both $y$ and $M_{ll}$ up to $1~{\rm TeV}$. Excluding the bins at the edge of the rapidity distributions, where statistical fluctuations are present, we see that the largest PDF uncertainty is at $y\sim 0$.
As explained in~\cite{Florian}, a study shows that NLO QCD corrections, applied on Monte Carlo (MC) and on PDF, enhance the cross section with respect to the LO prediction by $24-36\%$, with the largest NLO corrections at $y\sim 0$. A discrepancy of about $6\%$ is found between MRST-NLO and CTEQ-NLO PDF's.


\section{How to constrain PDF at LHC}\label{sec:PDFconstraints}

Several Standard Model processes are under study to constrain parton densities: the productions of $\gamma$, $W$ and $Z$ bosons and inclusive jets are equally important to constrain the parton densities and in particular the gluon density in complementary kinematic regions (see~\cite{Tricoli_Photon05}).

In G. Polesello's contribution we appreciate how the LHC jet data can be used to better constrain PDF fits: if the experimental systematic uncertainty is under control to $\leq 10\%$ level, LHC jet data can significantly contribute to constraining the high-$x$ gluon density with $1~fb^{-1}$ luminosity.  
Other studies~\cite{Hollins} have also shown that the prompt photon production process is extremely sensitive to PDF differences and can probe the perturbative theory of the gluon at high-$x$: the discrepancy between MRST2004-NLO, CTEQ6.1M and older PDF sets can be of the order of $16-18\%$ on the photon $\eta$ and $\pt$ distributions. \\
Furthermore, the $bg \rightarrow Zb$ process is sensitive to the $b$-quark content of the proton and the LHC predictions for the $Z+b$ cross-section, using different PDF sets, are $\pm 5-10\%$~\cite{Zb_VerducciEtAl}.



\subsection{W rapidity distributions}

A few days of LHC running at the nominal low luminosity ($10^{33} {\rm cm^{-2} s^{-1}}$) are sufficient to make the statistical uncertainty negligible with respect to the systematic uncertainties on $W$ cross section.
Among the systematic uncertainties there are experimental and theoretical contributions.


The ATLAS strategy for selecting $W$ bosons consist of identifying an isolated and highly energetic lepton, $E_T>25~{\rm GeV}$, and requiring a large amount of missing energy in the event due to the neutrino escaping detection, $\etmiss>25~{\rm GeV}$.  
The analysis of $W\rightarrow e \nu_e$ events fully simulated in the ATLAS detector, in the early data scenario, shows that the $W$ boson is a very clean signature: the trigger and the electron off-line identification with the electron $E_T$ and $\etmiss$ cuts leave a background contamination dominated by QCD events (less than $5\%$) and $W\rightarrow \tau \nu_\tau$ (about $0.5\%$). If a jet veto cut is added, the QCD background can be further reduced to a level of $\le 1\%$~\cite{CSCNote}.
Therefore the $W$ sector is an ideal environment to study and constrain theoretical and experimental systematics.

\subsubsection{Higher order corrections}
The differential cross section $d\sigma/dy$ for $W$ production has been calculated to the NNLO order in QCD with an energy scale uncertainty of  $\leq 1\%$~\cite{NNLO_WZ_y}.
With this level of precision in perturbative QCD calculations, the electro-weak (EW) contributions are no more negligible. As presented in this workshop, leading order electro-weak contributions with multi-photon radiation introduce corrections of the order of few percent on $W$ boson cross-sections.
The EW corrections, computed by the program HORACE interfaced to HERWIG in the $\alpha(0)$ scheme in the muon channel~\cite{CarloniPolesello}, are constant in rapidity and are about $3.5\%$ for a cut on the muon transverse momentum of $\pt>25~{\rm GeV}$ and can be up to $5.2\%$ for loser $\pt$ cuts. The dependence on the muon charge is negligible (up to $0.4\%$ for lose $\pt$ cuts)~\cite{TricoliCooper-Sarkar}.
Considering that these corrections in the muon channel are flat in rapidity and negligible on the muon-charge asymmetry, we can state that they do not have an impact on the PDF extraction, however they are relevant for luminosity measurements in order to achieve a precision of $6\%$ or better.
The electron channel needs further investigation. 

\subsubsection{PDF uncertainty on $W^{\pm}$ rapidity distribution.}

\begin{figure}[!tb]
 \hspace{-5mm} 
  \begin{minipage}[t]{.40\textwidth} 
  	\begin{center} 
       \includegraphics[width=7cm, height=6.5cm]{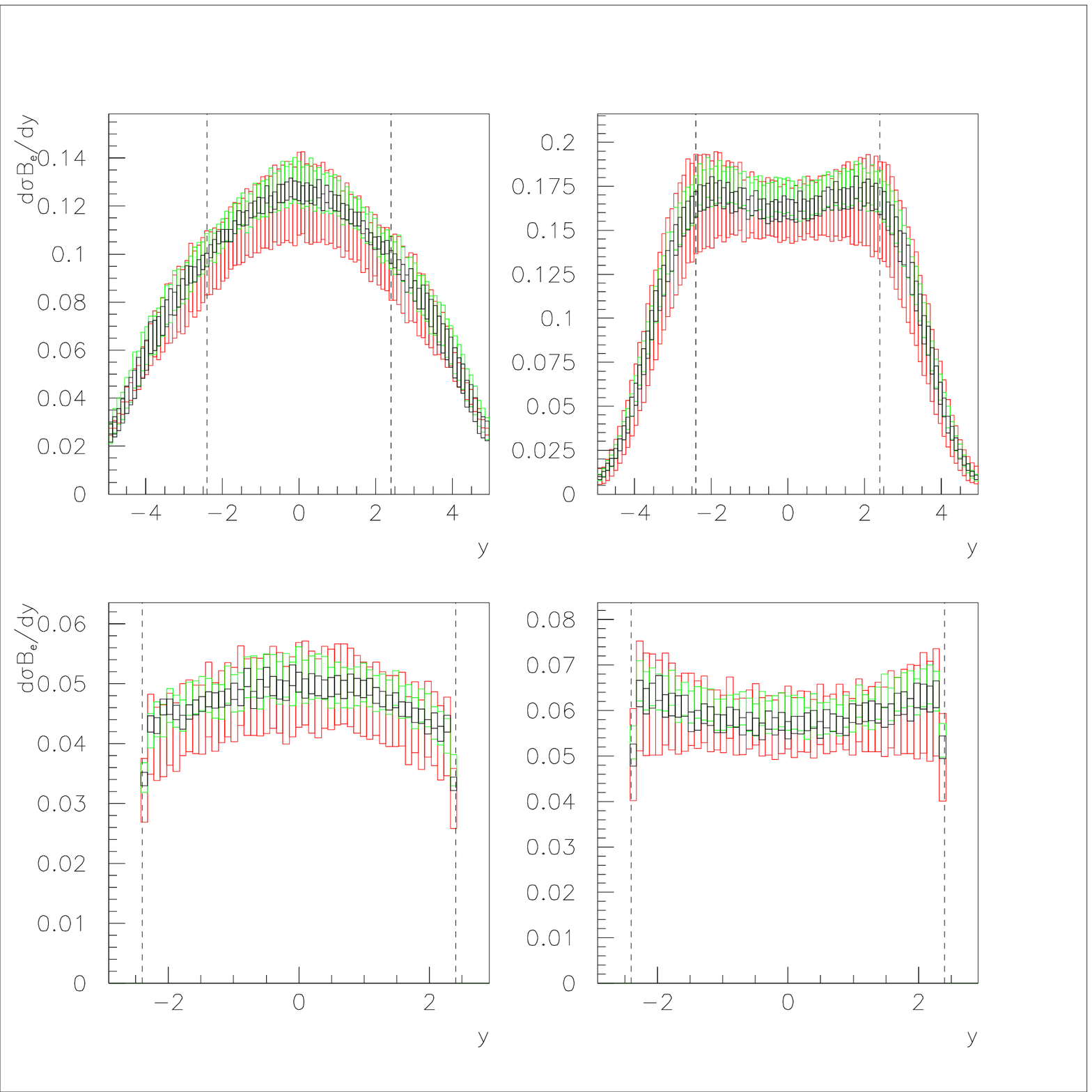}  
	\end{center}
  \end{minipage}
  \hspace{15mm}
  \begin{minipage}[t]{.40\textwidth}
    \begin{center}  
      \includegraphics[width=7cm, height=6.5cm]{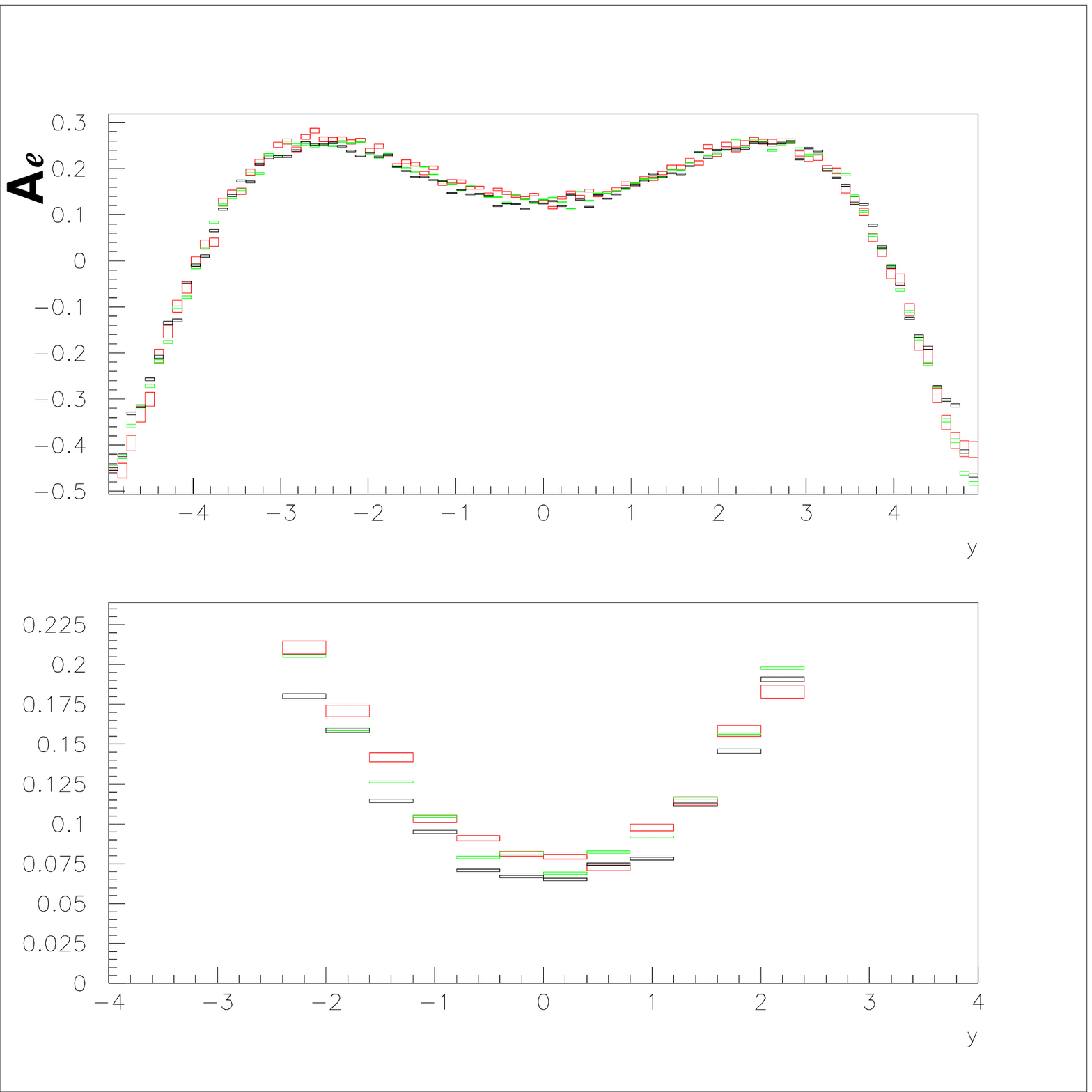}
    \end{center}
  \end{minipage}
  \caption[MC simulated $y$-spectra for $e^{\pm}$ and $A_l$ with PDF errors]{\label{fig:eplus_emin_rap_MC}}{\emph{HERWIG simulations of $e^{\pm}$ from $W^{\pm}$ decay, with CTEQ6.1M (red), MRST2001 (black) and ZEUS-S (green) PDF's and their quoted uncertainties (estimated with the PDF reweighting technique as in sec.~\ref{sec:PDFreweighting}). The top plots are at generator level, the bottom plots at ATLFAST detector level. Left fig: $e^-$ (left plots) and $e^+$ (right plots) rapidity spectra with NLO-QCD corrections.  Right fig: electron-charge asymmetry~\cite{Tricoli}.}}
\end{figure}

From fig.~\ref{fig:eplus_emin_rap_MC} we can see the full PDF uncertainties for three different PDF analyses, on the rapidity distribution of $e^{\pm}$ originating from $W^{\pm}$ decays. Their predictions are compatible within their uncertainties, which are in the range $4\%-12\%$, and are dominated by the gluon density. 

In a previous paper~\cite{HERAtoLHC} it is demonstrated that the LHC can improve the current constraint on the low-$x$ gluon parameter $\lambda_g$ ($xg(x)\approx x^{-\lambda_g}$) by more than $41\%$ by fitting the $e^+$ and $e^-$ rapidity distributions, if their experimental systematic uncertainties are kept under $5\%$ level.  

In the lepton-charge asymmetry $A_l=(\frac{d\sigma}{d\eta}^{l^+} -\frac{d\sigma}{d\eta}^{l^-} )/(\frac{d\sigma}{d\eta}^{l^+} + \frac{d\sigma}{d\eta}^{l^-})$ most of the gluon uncertainty cancel out leaving the valence up ($u_V$) and down ($d_V$) densities as main contributions to the total PDF uncertainty, which is reduced to $\sim 5\%$ at $\eta\approx 0$. However a discrepancy of $\sim 15\%$ is present at $\eta\approx 0$ between the MRST2002 and other two PDF's, CTEQ6.1M and ZEUS-S~\cite{Mandy}. In fact the MRST PDF's prediction for $u_V-d_V$ valence density is different from the other PDF's and is outside the quoted PDF uncertainty bands. This difference in current PDF fits comes from the lack of data on valence quantities at such low-$x$.
The LHC can be the first experiment to perform such measurement in the kinematic region $x\approx 10^{-3}$ and $Q^2=M_W^2$. 


\subsubsection{A posteriori inclusion of PDF's in NLO calculations.}
The MC computation of QCD final state observables to NLO is a lengthy process. In order to study the impact of PDF uncertainties on QCD cross section measurements in a faster way and allow for PDF fitting of these quantities, the technique of ``a posteriori'' inclusion of PDF's in NLO calculations has been developed for LHC processes~\cite{Carli-Salam-Siegert}~\cite{Clements-et-al}. A MC run is used to generate a grid (in $x_1$, $x_2$ and $Q$) of cross section weights that can subsequently be combined with an arbitrary PDF set. This enables the decoupling of the lengthy calculation of perturbative MC weights from the convolution with the parton densities.
Perturbative coefficients for jet (using NLOJET++), W and Z boson (using MCFM) production processes can be collected on a grid with an accuracy better than $0.02\%$.

\section{PDF reweighting of Monte Carlo events}\label{sec:PDFreweighting}
The computation of the full PDF uncertainty on a physics process is a cumbersome procedure. Given one PDF set, such as CTEQ or MRST, it requires the generation of twice as many MC samples as the number of free parameters in the global fit.
Furthermore one error analysis might not be sufficient since, as seen above, there can be large discrepancies between the results of different error analyses. 

A PDF reweighting technique has been studied and tested, requiring only one Monte Carlo generation with one conventional PDF set
~\footnote{This techniques is not as reliable if the PDF set is as ``unconventional'' as MRST2003, i.e. the validity of its kinematic space is smaller than the one available to the LHC. }~\cite{Tricoli}\cite{HERAtoLHC}.

This technique has been implemented using hard process parameters of the MC generation: flavours ($flav_{1}$ and $flav_{2}$) and  momentum fractions of the interacting partons $x_{flav_{1}}$, $x_{flav_{2}}$ and  the energy scale $Q$. The PDF set used for the MC generation is named $PDF_{1}$.


The PDF reweighting technique consists of evaluating, on the event-by-event basis, the probability of picking up the same flavoured partons with the same momentum fractions $x_{flav_{1}}$, $x_{flav_{2}}$, according to a second PDF set, $PDF_{2}$, at the same energy scale $Q$
, then evaluating the following ratio

\begin{equation}
Event~Weight = \frac{f_{PDF_{2}}(x_{flav_{1}},Q)}{f_{PDF_{1}}(x_{flav_{1}},Q)}\cdot 
\frac{f_{PDF_{2}}(x_{flav_{2}},Q)}{f_{PDF_{1}}(x_{flav_{2}},Q)}~.
\end{equation} 

After the $Event~Weight$ is applied on MC events generated with $PDF_{1}$, they will effectively be distributed according to $PDF_{2}$. 

This technique has been tested using HERWIG (for inclusive W production) and ALPGEN interfaced to HERWIG (for W+jets production) as Monte Carlo generators and with various recent PDF sets. Similar results have been obtained with these two MC generators and with different PDF sets, as discussed below.
\begin{figure}[!tb]
  \begin{minipage}[t]{.40\textwidth} 
  	\begin{center} 
       \includegraphics[width=7.0cm, height=6.5cm]{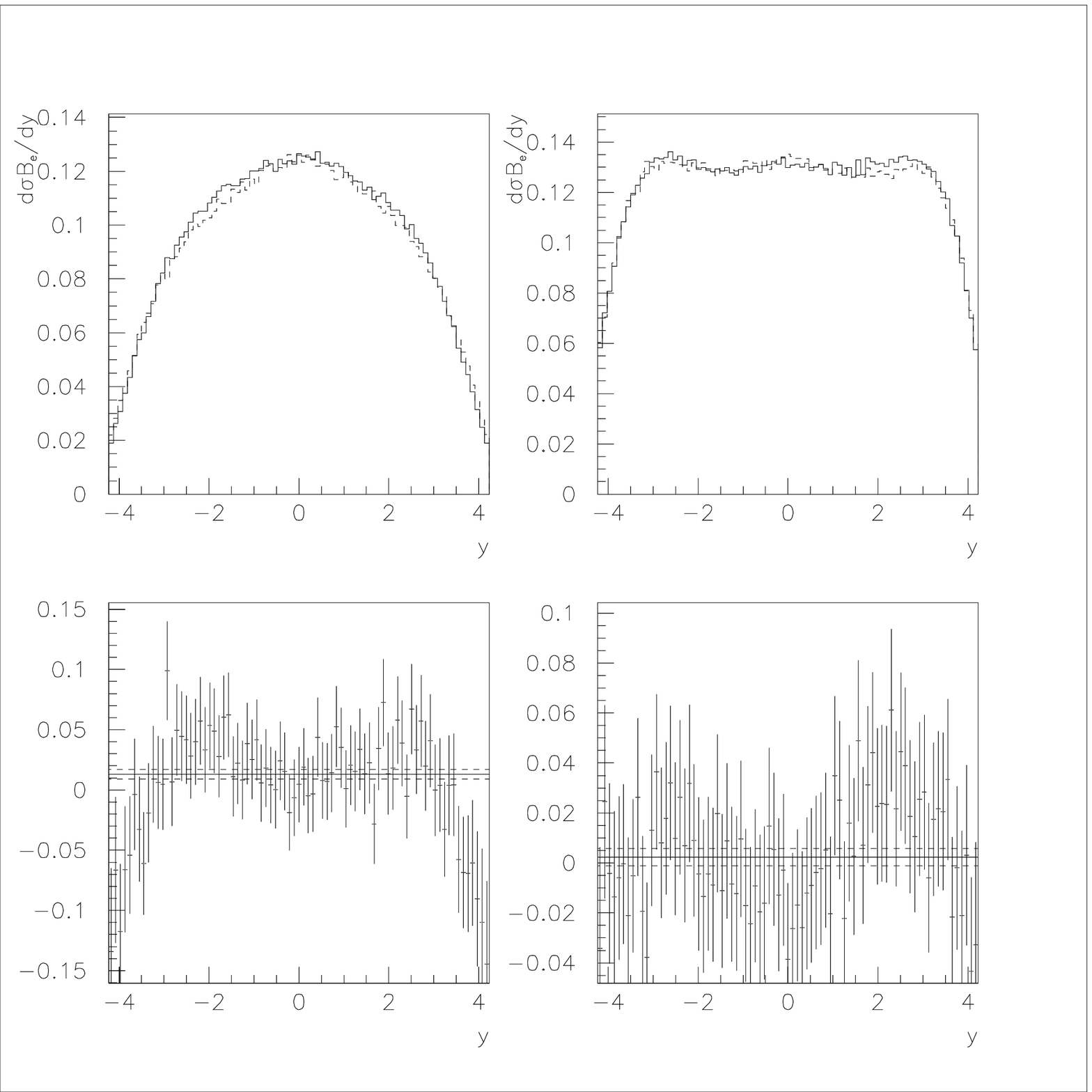}  
	\end{center}
  \end{minipage}
  \hspace{15mm}
  \begin{minipage}[t]{.40\textwidth}
    \begin{center}  
      \includegraphics[width=7.0cm, height=6.5cm]{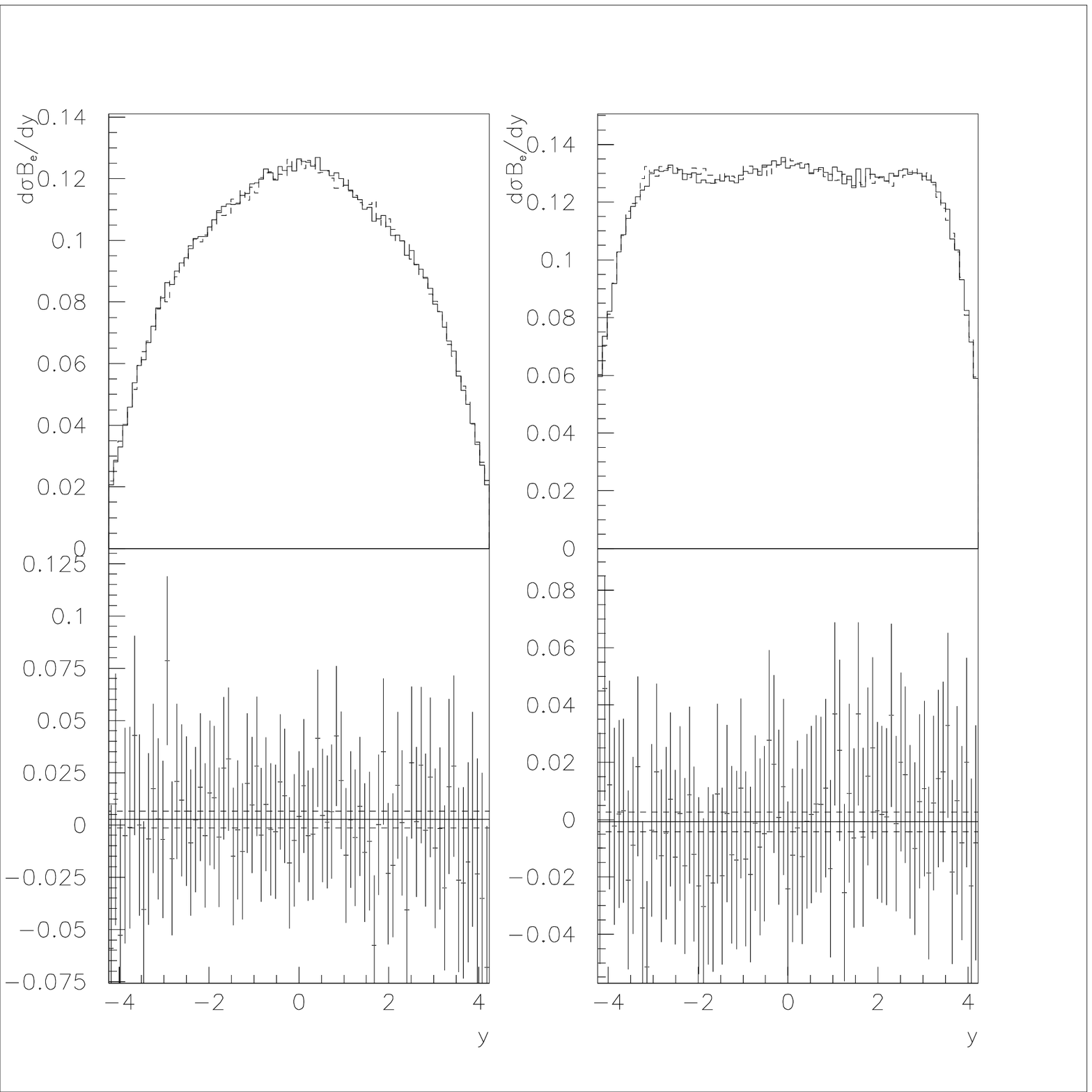}
    \end{center}
  \end{minipage}
  \caption[PDF reweighting]{\label{fig:PDFreweighting}}{\emph{Left fig: $W^{-}$ and $W^{+}$ rapidity distributions at HERWIG generator level for events generated with CTEQ6.1M (dashed lines) and for events generated with MRST2002 (solid lines) and their relative differences (at the bottom).
The straight lines are the means of the points with uncertainty bands. Right fig: same as left hand side plots for events generated with CTEQ6.1M (dashed lines) and for events generated with MRST2002 and PDF-reweighted with CTEQ6.1 (solid lines). Similar results have been obtained reweighting between MRST2002 and ZEUS-S PDF's.}}  
\end{figure}
Fig.~\ref{fig:PDFreweighting} shows the accuracy of this technique using HERWIG: the bias over the all $y$ range is of the order of $0.5\%$ or less and there is no evidence of $y$ dependence. 
Comparing the bottom plots on the right and left hand sides of fig.~\ref{fig:PDFreweighting} we see that the PDF reweighting technique corrects for the difference in normalisation between $PDF_1$ and $PDF_2$
 and corrects for the $y$ modulation.

This technique can be used to estimate the full PDF uncertainty, starting from one sample of MC generated events, for distributions that are determined by the MC hard process.

%
\bibliographystyle{atlasstylem}
\bibliography{%
atlaspaper}

%
%
\newpage

\end{document}